\documentclass[prl,twocolumn,showpacs,showkeys,preprintnumbers,amsmath,amssymb]{revtex4}

 \usepackage{epsf}

\usepackage{amsmath,amsfonts}

\newcommand{\tr}{\hbox{ Tr}}

\begin{document}

\title{Possible exotic stringy signatures at the LHC}
\author{David Berenstein}
\email{dberens@physics.ucsb.edu}


 \affiliation{Department of
Physics, University of California
at Santa Barbara, CA 93106}

\begin{abstract}
In this paper I discuss some string inspired exotic colored matter particles that could be produced at the LHC, as well as the experimental signature for the observation of such states. 
Their most easily identifiable signature of this scenario is that many of these exotic particles would decay into standard model particles only via higher dimension operators in the effective field theory at a TeV scale. Some of these particles would only have three body decays into  standard model particles.
\end{abstract}
 \maketitle
\begin{center}{\em Introduction}
\end{center}
In the near future, experiments in high energy physics at the LHC will be able to probe a new regime of energy scales for particle production.  The  fact that we do not control the initial state precisely (as any process involves the hard scattering of partons with random fractions of the total momentum of the proton)  makes it a  lot harder to understand what the new physical states could be. Moreover, these new particles
could carry color quantum numbers, and therefore if they exist, they will be produced in pairs copiously by gluon fusion processes. If the new particles decay into standard model particles, they might be missed because of background events, or just because one is not looking in the right channel. 

A particularly attractive scenario for LHC is the possibility that a fundamental theory of interactions that include gravity might be just around the corner. Such a scenario would involve large extra dimensions \cite{ADD}. The current best guess for a fundamental theory with such properties is the idea of string theory. A string model can usually be characterized in terms of a string scale $\ell_s\sim M^{-1}$, which appears below the Planck scale. The string scale could be as low as a few TeV, and therefore one would expect to see a few string states produced at the LHC. However, one would not immediately expect to see the typical Regge trajectories associated to a string model, because the reach at the LHC is finite. The most important property of these models is that although they are unlikely (this reflects our theoretical prejudice), they have not been ruled out yet and they could be experimentally accessible in the near future. For comparison, current bounds from LEP give lower estimates for direct searches at around 100 GeV for new charged particles beyond the standard model  \cite{Abb} and one would expect that in a string model various new particles would be charged.

The question I will address in this paper is how the physics beyond the standard model for these scenarios could look like. There is in general quite a bit of model dependence associated to this question. However, if one tries to understand the problem from a bottom-up approach where one fits the standard model in some hypothetical brane construction and one asks what else could be seen, one finds that the spectrum of particles that would be accessible is limited and the possibilities can be enumerated.  Moreover, the quantum numbers  of the new particles can 
be inferred by general considerations. If one demands that the string model is perturbative (after all, the standard model is well described by perturbation theory at intermediate scales),  the low energy string scale forces us to consider D-brane  setups in order to be compatible with large extra dimensions.
In this scenario, perturbative string theory considerations would
dictate the approximate strength of various coupling constants, so that one could have a relatively clear signal pointing to a stringy origin of dynamics.

\begin{center}
\em Perturbative D-brane models
\end{center}

The concept of D-branes was introduced in \cite{DLP}. 
They are described as a geometric locus where strings can end. In a theory of quantum gravity like string theory, any defect or extended object in spacetime can bend and it will therefore have excitations. The excitations of D-branes are the open string attached to them. Moreover, one can have various D-branes on top of one another. In these situations one finds that one needs to consider open strings with Chan-Paton indices on them, and one has continuous gauge symmetries associated to the ends of the string. These constructions can lead to a model of particle physics on a brane and a possible realization of the standard model within string theory.

The allowed gauge groups are those that can have a large $N$ limit: $U(N), SO(N), Sp(N)$.
Also each end of the string ends carries fundamental charge with respect to the stack of branes on which it ends. Thus any open string will carry the quantum numbers associated to some type of bifundamental representation. Typical models consider various branes intersecting each other at angles\cite{BDL} and as calculated in \cite{DLP}, once one considers unoriented models, there is no a-priori restriction on the 2-index tensor  representation of the gauge groups that the strings carry. A recent review of such top-down constructions can be found in \cite{BCLS}.

Finally, in the perturbative regime where the low energy dynamics is given in terms of open strings alone (all other non-perturbative states are heavy), the interactions are generated by disc diagrams.
These are single traces of fields. If a vertex has $n+2$ particles attached to it, it will appear with a coupling constant dependence of $g^{n}$,  where $g$ is the open string coupling constant. The anomaly cancellation condition also has to be imposed: it is a consistency condition of the brane model (in explicit models this is a subset of  the conditions imposed by tadpole cancellation). The presence of $U(1)$ fields allows for the possibility of mixed anomalies. These are cancelled by the Green-Schwartz mechanism. In four dimensions, this requires some type of pion (or axion) from the closed string sector to carry the anomaly and to transform inhomogeneously under gauge transformations. This amounts to a Stueckelberg mass for the corresponding gauge boson, where the pion is eaten up by the corresponding gauge 
field. This produces massive $Z'$ vector fields and serves as a mechanism to avoid proton decay \cite{IQ}. 

It is convenient to introduce a graphic notation to describe the field theory degrees of freedom. These are called quivers or moose diagrams. The gauge degrees of freedom (brane stacks) will be described by nodes in a graph. The open string particles will be given by edges connecting two vertices and arrows that dictate if the corresponding end of the string is fundamental or
anti-fundamental (there is no such distinction for $SO$ or $Sp$ groups, so one should not add arrows at those ends). 
One should also label the edges according to the other quantum numbers that the particles carry.

In an effective field theory approach, one writes a lagrangian with all  the particles that one can access
below a given scale, and all possible operators compatible with the single trace condition and the symmetries of the system. These will have coefficients of order one multiplying the appropriate power of the string coupling constant, and
dimensionally corrected by powers of $\ell_s^{-1}$ for non-renormalizable couplings. 
This is the same philosophy espoused in \cite{BJL}.
One can also show that the stack of branes can not be unified
as many Yukawa couplings vanish \cite{CPS}. Moreover, right handed quarks can not be in a two index antisymmetric representation of $U(3)$ \cite{Bguts}. 

\begin{center}
\em Exotic particles from  fusion of known particles
\end{center}

The main motivation for writing this paper is to understand how stringy dynamics at a few TeV
might be discerned at the LHC. This means that we must relax the ideas of taking just the low energy/massless limit and we have to look for a more comprehensive understanding of the effective dynamics below a few TeV.  We will look for the possibility of 
particles with exotic quantum numbers that might be relevant in the LHC experiment and how
to deduce their quantum numbers.

The first observation we want to make is that typically the open string states have various 
spins and masses, but that if the spin is higher than one, then these states have a mass higher than the string scale (which we are assuming is slightly beyond reach). This is most easily seen on configurations of strings at angles 
\cite{BDL}. The simple observation is that one needs vertex operators that are roughly of the form $:\partial X^\mu\partial X^\nu \exp(ikX)\xi :$ where $\xi$ is an open string ground state. In unitary Conformal Field Theories the conformal dimension of $\xi$ is greater than zero, so the mass associated to this particle is greater than one. One can do a similar argument if $\xi$ is a ground state in the Ramond sector, so that it carries spin quantum numbers. For more information  see \cite{Pol}.

This sketched argument shows that the spin of states below the string scale will be given by spin $0, 1/2, 1$ states, giving rise to a spectrum  of ordinary particle physics,  with 
non-renormalizable interactions characterized by the string scale and powers of the string coupling constant.

Based on these facts, first one finds an embedding of the standard matter content into a gauge theory and require the minimal amount of extra matter to complete the multiplets and satisfy the abstract D-brane consistency of the model. One does not require supersymmetry.
Then one can deduce what additional matter content is allowed below the string scale. 
I will assume for simplicity that any particle that one adds at this stage is non-chiral (fermions have Dirac masses) and that all vector particles below the string scale have been accounted for by the gauge group.

\begin{figure}[ht]
\begin{center}
\epsfysize=2.0 cm\epsffile{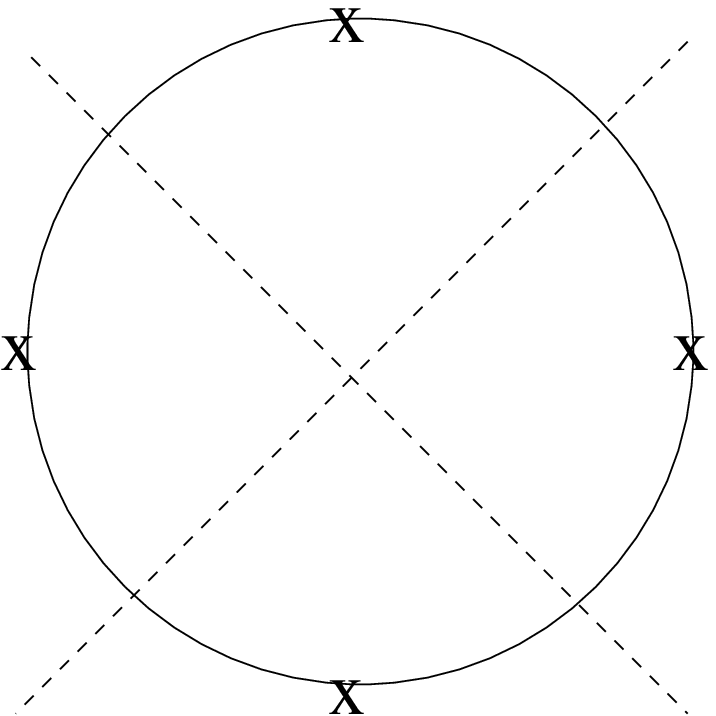}
\end{center}\caption{A hypothetical interaction in the lagrangian with four fields. The cuts indicate intermediate states in the s and t channels}\label{fig:cut.eps}
\end{figure}

One can use factorization of $n+2$-point disc vertices to deduce the possible quantum numbers of string states that are not know already by cutting disc diagrams as in figure \ref{fig:cut.eps}. Cutting the figure along a channel, there will be off-shell string  states propagating along the cut.

 As I have argued, we want to project the new states into the spin zero and spin one-half multiplets (those with low mass), and to ignore the higher spin states. The cuts obtained this way show that new states will be obtained by gluing together the fields that are already known. There can be various particles with the same spin, but different masses in a given channel. 

The rules for combining the states are that one follows paths in the quiver diagram being careful that arrows are multiplied so that two consecutive edges have arrows into and out of a node. This can be done with any higher point function, cutting the disc along any preferred path. Factorization gives us
also the couplings (Feynman rules) of the new particles.
 
In order to make this approach concrete, I will show how this works in one particular extension of the standard model that is compatible with the rules of low energy D-brane dynamics \cite{BP}. This is chosen because there is only one particle beyond the standard model that is required to make an embedding of the standard model into a D-brane possible. 
The extra particle is a massive Z' that couples mostly as baryon number, and corresponds to the fact that in D-brane models the gauge group on a D-brane is never just $SU(N)$, but $U(N)$. Generic D-brane models usually predict many $Z'$ particles that mix in a complicated way \cite{GIIQ}. 

\begin{center}
\em Extensions of the MQSM and their LHC signatures
\end{center}

\begin{figure}[ht]
\begin{center}
\epsfysize=3.5 cm \epsffile{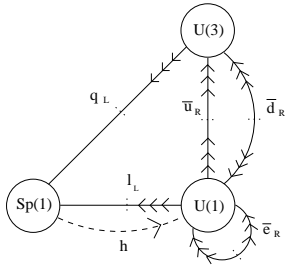}
\label{fig:mqsmquiv.eps}
\end{center}
\caption{The quiver representation of the MQSM. Only left chiral fermion fields are shown. The Higgs doublet $h$ is shown with a dashed line. The complex conjugate field $\bar h$ is not shown, but can be inferred from the diagram. The group shown as $Sp(1)$ is the same as $SU(2)$}
\end{figure}

The minimal quiver embedding of the standard model into a D-brane like setup was described in \cite{BP} and was called the Minimal Quiver Standard Model (MQSM). The model is given by the following quiver diagram \ref{fig:mqsmquiv.eps}. 

Now, consider other states of spin less than one that can be added to the MQSM by gluing arrows. For example, if one follows the path $h \ell$, one might end up with a 
$SU(2)_W$ triplet or singlet, and with spin one half. These would be akin to right handed 
neutrinos or gauginos. 

The most interesting cases involve new colored particles. This is because they will be produced in large numbers at the LHC by the process of gluon fusion. One can easily find paths that would lead to a fourth generation of Dirac quarks, and they could be doublets or singlets under $SU(2)_W$. One can also show that since the Higgs and the leptons have the same gauge quantum numbers, any new composite arrow in the diagram can be either a boson or a fermion.  

The more interesting fermion exotics would not be charged under the weak interaction and would have two indices in the $U(3)$ color. They can come in three varieties: adjoints (in which case they would split into gauginos and an extra neutral particle under the SM gauge group), or one could have symmetric and antisymmetric tensors from the
$\bar u e_R \bar u$ path and it's complex conjugate. 
These would carry baryon number $2/3$. They would also carry electric charge equal to 
$1/3$.  We present these states in  table \ref{tab:2t}
\begin{table}[ht]
\begin{tabular}{|c|c|c|c|c|}
 \hline
Field & $SU(3)_c$ representation & $U(1)_B$ & Q \\
\hline $\psi^{[ab]}\!_\alpha$ & ${\bf 6}$ & $2/3$ & 1/3 \\
 $\psi^{(ab)}\!_\alpha$ & ${\bf \bar 3}$ & $2/3$ & 1/3 \\
 $\psi^a_{b \alpha}$ & $\bf 8\oplus 1$ & $0$ & 0 \\
 \hline
\end{tabular}\caption{Possible fermionic stringy particles with 2 tensor representations of $SU(3)_c$ in extensions the MQSM. The color indices are $ab$, the left Weyl spinor index is $\alpha$. The complex conjugate representations are also allowed, and should be written with a bar on the field. The adjoint representation is real.}\label{tab:2t}
\end{table}
If any such particle is produced, it will likely decay into standard model particles. This is going to be controlled by writing the most general terms in the effective field theory lagrangian compatible with all the gauge symmetries.

For example, consider the spin $1/2$ particle $\psi^{ab}$(in the symmetric or antisymmetric representation of $U(3)$). It has baryon number $2/3$ and carries charge equal to one third. The lowest order term in the lagrangian that has only one $\psi$ and standard model particles necessarily involves two quarks, and for making a Lorentz invariant term, one needs an extra term from a lepton. There are two  classes of terms in the lagrangian at dimension 6
\begin{equation}
 \frac{ e g_s  \psi^{ab} }{M^2}\left((\bar u_R)_a(\bar u_R)_b \bar e_R \xi_1
+ (\bar q_L)_a \cdot (\bar\ell_L) \bar u_R  \xi_2\right)
\end{equation}
These produce three body decays $\psi \to u+u+e$, or $\psi \to u+ d+ \nu$.
If $\psi$ is the lightest Beyond the Standard Model  particle, there will be no intermediate two body decays. Similar observations were made in \cite{Kang}.
The parameters $\xi_1$ and $\xi_2$ are matrices that depend on flavors. There are also various Fierz rearrangements of the first term that are possible. 
The lifetime will be of order 
\begin{equation}
\Gamma\sim |\xi|^2 \alpha_{em} \alpha_s \frac {m_\psi^4}{M^4} m_\psi\sim 10^{-3} \frac {m_\psi^4}{M^4} m_\psi
\end{equation}
where any numerical factors can be absorbed in the definition of $\xi$. These states are narrow relative to their mass and can be very long lived if $m/M$ is sufficiently small. 

The most visible channel for detection would be in events with four jets and opposite sign leptons. The jets and leptons will have large transverse energy, and a sufficiently high threshold would reduce Standard Model background contributions. They will also be central in the detector since the $\psi \bar \psi$ pairs are produced at threshold and moving relatively slowly. These events will have no missing energy, and one could have complete reconstruction of the mass from combining the four-momenta of two jets and a lepton for the two different decays within each event. One should see a large excess of events with coinciding masses reconstructed at the $\psi$ mass. 

If these fermions are accidentally much lighter than the string scale (let us say $m_\psi \sim 500 GeV$ and $M\sim 10^4 TeV$), they would be long-lived enough to show displaced 
vertices and they would also hadronize giving rise to a rich spectrum of bound states.
Such a possibility was suggested in \cite{WZ}, where at the renormalizable level these extra particles in the ${\bf 6, \bar 3}$ are stable. A study for the discovery of heavy hadrons was carried in \cite{KHN}, where it is assumed that the hadrons are long lived enough to make it to the calorimeter. One would consider these models as finely tuned, but one could argue 
on the same principles as was done for split supersymmetry that this possibility should not be discarded \cite{AHD}.

For completeness, one should also study possible scalar particles. The most interesting situations are those that do not produce decays via renormalizable interactions. Indeed, for renormalizable interactions of colored scalar particles, one usually expects to have Yukawa couplings involving the quarks and possibly the leptons. These tend to give large  
contributions to Flavor Changing Neutral Current (FCNC) processes from box diagrams, which are very constrained \cite{PDG}. Without tuning their interactions, their masses tend to be pushed to a larger scale. However, if one only has dimension five or higher operators available, loop integrals for FCNC processes (being associated to high dimension operators) will be determined by the cutoff scale $M$, rather than the mass of the scalar particle itself.

Of these scalars, the adjoints of $U(3)$ are interesting. They would likely decay to two gluons 
via operators $\tr(s F F)$, or also to two quarks via $\tr(D_\mu s  \bar q_L \sigma^\mu q_L)$ and a similar term for the right handed quarks. Seemingly suppressed by an extra power of $g$, one would also have an operator $\tr(s q h \bar u)$ that could represent decays into two or three particles (once the higgs gets a vev).
Notice that the two operators describing decays with two quarks reflect the same physics: one can integrate by parts the derivative and use the equations of motion of the spinors to produce the second operator. Thus the fist term produces decays suppressed by an extra power of $g$ relative to a naive counting. This is because in the massless limit for quarks the bilinear current would be approximately conserved. This is an example of redundant couplings (for more details see \cite{W}, section 7.7). This means that the decay widths into two and three bodies involving fermions are roughly of the same order of magnitude.

Even more dramatic is the case of a scalar 
partner of the $\bar u_R$ quarks: only dimension five or higher operators with one such scalar would be allowed and their decay would necessarily involve at least one quark and one charged lepton. A three particle decay has to involve a Higgs or a gauge boson plus two fermions at roughly the same rate.

If realized in nature, these scalars could provide a clean factory for Higgs production. 
Notice that in the standard model the Higgs decays most of the time into $b\bar b$ jets. In  standard model processes there is a huge background of such events. However, in the case being discussed here, the combinatorics of reconstructing the mass of the scalar particle $s$ would reduce the background considerably and one should be able to read the mass of the Higgs from the b jets. This leads to the exciting possibility of being able to test directly the coupling of the Higgs particle to the masses of various elementary particles at the LHC.

{\em Acknowledgements:} The author would like to thank C. Campagnari, M. Einhorn, D. Stuart for discussions related to this paper. Work supported in part by 
 the U.S. Department of Energy, under grant DE-FG02-91ER40618.


\begin{thebibliography}{99}



\bibitem{ADD}
  N.~Arkani-Hamed, S.~Dimopoulos and G.~R.~Dvali,
  Phys.\ Lett.\  B {\bf 429}, 263 (1998)
  [arXiv:hep-ph/9803315].
  I.~Antoniadis, N.~Arkani-Hamed, S.~Dimopoulos and G.~R.~Dvali,
  Phys.\ Lett.\  B {\bf 436}, 257 (1998)
  [arXiv:hep-ph/9804398].
 
\bibitem{Abb}
  G.~Abbiendi {\it et al.}  [OPAL Collaboration],
  Eur.\ Phys.\ J.\  C {\bf 35}, 1 (2004)
  [arXiv:hep-ex/0401026].
  
\bibitem{DLP}
  J.~Dai, R.~G.~Leigh and J.~Polchinski,
  Mod.\ Phys.\ Lett.\  A {\bf 4}, 2073 (1989).
  
\bibitem{BDL}
  M.~Berkooz, M.~R.~Douglas and R.~G.~Leigh,
  Nucl.\ Phys.\  B {\bf 480}, 265 (1996)
  [arXiv:hep-th/9606139].
  
\bibitem{BCLS}
  R.~Blumenhagen, M.~Cvetic, P.~Langacker and G.~Shiu,
  Ann.\ Rev.\ Nucl.\ Part.\ Sci.\  {\bf 55}, 71 (2005)
  [arXiv:hep-th/0502005].

\bibitem{IQ}
  L.~E.~Ibanez and F.~Quevedo,
  JHEP {\bf 9910}, 001 (1999)
  [arXiv:hep-ph/9908305].


\bibitem{BJL}
  G.~Aldazabal, L.~E.~Ibanez, F.~Quevedo and A.~M.~Uranga,
  JHEP {\bf 0008}, 002 (2000)
  [arXiv:hep-th/0005067].
  D.~Berenstein, V.~Jejjala and R.~G.~Leigh,
  Phys.\ Rev.\ Lett.\  {\bf 88}, 071602 (2002)
  [arXiv:hep-ph/0105042].



\bibitem{CPS}
  M.~Cvetic, I.~Papadimitriou and G.~Shiu,
  Nucl.\ Phys.\  B {\bf 659}, 193 (2003)
  [Erratum-ibid.\  B {\bf 696}, 298 (2004)]
  [arXiv:hep-th/0212177].


\bibitem{Bguts}
  D.~Berenstein,
  arXiv:hep-th/0603103.

\bibitem{Pol}
  J.~Polchinski,
{\it  Cambridge, UK: Univ. Pr. (1998) 531 p}

\bibitem{BP}
  D.~Berenstein and S.~Pinansky,
  Phys.\ Rev.\  D {\bf 75}, 095009 (2007)
  [arXiv:hep-th/0610104].

\bibitem{GIIQ}
  D.~M.~Ghilencea, L.~E.~Ibanez, N.~Irges and F.~Quevedo,
  JHEP {\bf 0208}, 016 (2002)
  [arXiv:hep-ph/0205083].
 
 
\bibitem{Kang}
  J.~Kang, P.~Langacker and B.~D.~Nelson,
  Phys.\ Rev.\  D {\bf 77}, 035003 (2008)
  [arXiv:0708.2701 [hep-ph]].
 
\bibitem{WZ}
  F.~Wilczek and A.~Zee,
  Phys.\ Rev.\  D {\bf 16}, 860 (1977).
 
\bibitem{KHN}
  A.~C.~Kraan, J.~B.~Hansen and P.~Nevski,
  Eur.\ Phys.\ J.\  C {\bf 49}, 623 (2007)
  [arXiv:hep-ex/0511014].
  
\bibitem{AHD}
  N.~Arkani-Hamed and S.~Dimopoulos,
  JHEP {\bf 0506}, 073 (2005)
  [arXiv:hep-th/0405159].
 
\bibitem{PDG}
  W.~M.~Yao {\it et al.}  [Particle Data Group],
  J.\ Phys.\ G {\bf 33}, 1 (2006).
\bibitem{W}
  S.~Weinberg,
  ``The Quantum theory of fields. Vol. 1: Foundations,''
{\it  Cambridge, UK: Univ. Pr. (1995) 609 p}


\end{thebibliography}
\end{document}